# Leveraging Elastic instabilities for Amplified Performance: spine-inspired high-speed and high-force soft robots

Yichao Tang[1,2], Yinding Chi[2], Jiefeng Sun[3], Tzu-Hao Huang[4], Omid H. Maghsoudi[5], Andrew Spence[5], Jianguo Zhao[3], Hao Su[4], Jie Yin[1,2]*

**Affiliations**
[1]Department of Mechanical Engineering, Temple University, 1947 North 12[th] Street, Philadelphia, PA 19122 USA.
[2]Department of Mechanical and Aerospace Engineering, North Carolina State University, Raleigh, NC 27695 USA.
[3]Department of Mechanical Engineering, Colorado State University, Fort Collins, CO 80523 USA.
[4]Department of Mechanical Engineering, The City University of New York, New York, NY 10031 USA.
[5]Department of Bioengineering, Temple University, 1947 North 12[th] Street, Philadelphia, PA 19122 USA.

* Corresponding author. Email: jyin8@ncsu.edu

## Abstract

Soft machines typically exhibit slow locomotion speed and low manipulation strength due to the intrinsic limitation of soft materials. Here we present a generic design principle that harnesses mechanical instability for a variety of spine-inspired fast and strong soft machines. Disparate from most current soft robots that are designed as inherently and unimodally stable, our design leverages tunable snap-through bistability to fully explore the ability of soft robots to rapidly store and release energy within tens of milliseconds. We demonstrate the embodiment of the generic design principle with three high-performance soft machines, including high-speed cheetah-like galloping crawlers with locomotion speeds of 2. 68 body-length/s, high-speed underwater swimmers (0.78 body-length/s), and tunable low-to-high-force soft grippers with over $1\text{-}10^3$ stiffness modulation (maximum load capacity is 11.4 kg). Our study establishes a new generic design paradigm of next-generation high-performance soft robots that are applicable for multifunctionality, different actuation methods and materials at multiscales.

## Summary
Bistable spined soft robots enable high-speed cheetah-like galloping and fast-speed swimming, as well as high-force manipulation.

# INTRODUCTION

Recently, soft robotics have attracted tremendous research interest due to their safe and adaptive interaction with humans and harsh environments, enabling a wide range of new functionalities that can rarely be achieved by conventional rigid robots (*1*), including manipulation of delicate objects (*2, 3*), navigation through a confined space (*4, 5*), and actuation with multiple degrees of freedom (*6, 7*). Despite these advances, it remains challenging to achieve high performance in soft robots, e.g. high-speed locomotion and high-strength manipulation due to the intrinsic limitations of their soft bodies (*8*). Normally, high-speed locomotion requires fast response, large force output, high strain energy storage, and high precision motion of the actuators. Soft-bodied robots have typically shown small force exertion, large deformation, and slow response time, resulting from their materials softness and structural compliance. Soft materials make it difficult to rapidly store and release large amounts of mechanical energy, limiting their applications in high-speed locomotion and high-strength object manipulation (*9*). In fact, most studied soft robots to date demonstrate a relatively slow locomotion speed in the range of 0.02–0.8 body length/second (BL/s) both on land and underwater, as well as limited payload capability of grasping (*10*) in soft manipulators, which are far below the locomotion speed (1-100 BL/s) and high-strength object manipulation (> 10 kg) in animals and rigid robots. In addition, soft machines are often designed for single-purpose functionality in either locomotion or manipulation, a generic soft machine which serves multiple purposes under different working scenarios could largely simplify the design, as well as improve the working efficiency and its versatility.

In nature, a number of quadrupeds exhibit exceptionally high-speed locomotion that has evolved through millions of years of evolution. Locomotion principles have been garnered from their study and embedded into bioinspired terrestrial robots (*11*). Among the locomotor gaits, galloping is typically used at the highest running speeds, e.g., the fastest galloping cheetah demonstrates a record speed of 29 m/s on land. Galloping has also been suggested as the most energy-efficient mode of locomotion in quadrupedal mammals (*12*). The roles and benefits of spine flexion and extension for high-speed locomotion are investigated for vertebrate animals in terms of self-stabilization and elastic energy storage (*13*). Recently, active spine mechanisms (*14*) were studied in legged robots for high-speed and energetically economic locomotion (*15, 16*). It was found that active spine actuation can increase speed and enhance stability. However, the potential benefit of the active spine mechanism in soft locomotive robotics remains unexplored.

Most current soft-bodied robots are designed to be inherently stable and usually unimodally stable. Relaxing this design constraint allows exploration of the principle of elastic instability for rapid energy storage and release in a soft robot. Very recently, bistability has been studied in soft robotics for untethered directional propulsion (*17*), autonomous control of airflow in soft bistable valves (*18*), and soft fluidic actuators with amplified responses (*19*). Although bistable soft actuators have been applied to high power density design for hydrodynamic interaction (*20*), exploration and analysis of rigid and soft materials (*21*), and multi-linkage and multiple spring mechanisms (*22*), they have primarily been tailored for high-speed jumping locomotion.

To address these challenges, we present a generic design principle that harnesses mechanical instability for both high-performance and versatile soft robots ranging from high-speed terrestrial and aquatic locomotion to high-force manipulation. Inspired by the spine flexion and extension in high-speed yet energy-efficient quadrupedal mammals such as galloping cheetahs, we propose here a soft actuator with a tunable bistable spine mechanism for rapid energy storage/release, as schematically shown in Fig. 1A. Bistability enables both amplified actuation speed and forces for high-speed locomotion and wide-range stiffness modulation for high-force manipulation in soft actuators. Our design leverages high nonlinearity phenomena of instability to fully explore the ability of soft machines to rapidly store and release energy. Our work represents a new family of versatile actuators that are soft but strong with high-speed or high-force output and a wide-range of tunable stiffness, significantly expanding the spectrum of conventional ones made of rigid bodies and elastic joints (e.g., series elastic actuators) (*23*).

The bistable hybrid soft actuator presented here is constructed by combining spring-based bistable linkages as a "skeletal spine" and soft pneumatic bending actuators as "skeletal muscle" (Fig. 1B). It harnesses a rapid, switchable snap-through instability between two stable states that are actuated by soft pneumatic bending actuators. The mechanical energy of our actuator is stored in a linear spring connected to the linkage-based spine, which can be tuned from low to high energy-storage capacity through simple spring pretension or springs with different stiffness. This spring in soft actuators acts as a force amplifier. It significantly enlarges not only the actuation speed and force, but also the design space for stiffness modulation through tunable snap-through instabilities. We first study the design principle of a bistable hybrid soft actuator and characterize its actuation response. We develop an energy-based theoretical model to understand its nonlinear behavior, including the relationship among design parameters (e.g., spring stiffness and pretension length), control inputs (pneumatic actuation pressure), and outputs (including force, bending angle, and energy barrier). We then examine its amplified mechanical performance in both and response time and force output through bistability. Finally, we explore the simplicity of employing the tunable bistable mechanism for applications ranging from high-speed terrestrial and aquatic locomotion to high-force grasping.

## RESULTS

### 1. Design Principles of a Bistable Hybrid Soft Actuator

We leverage the physical principle of elastic instability to guide the design of fast and strong soft machines, where energy competition and synergy relationship between springs and soft pneumatic actuators determine the bistability of the hybrid soft actuator. Bistability provides two distinct dynamic operating regimes: 1) bistable mode switching between two stable states (path (a) in Fig. 1C). Upon activation, highly-nonlinear, reversible, snap-through bistability triggers instantaneous and significant changes in speed and position through rapid energy storage and release (Fig. 1C), which guides the design of high-speed soft crawler and swimmer. 2) Monostable mode bending beyond the equilibrium configuration (path (b) in Fig. 1C). The tunable pretension in the spring provides high-force output with a wide range of stiffness modulation for the design of high-force soft grippers.

To understand the design methodology and relationship between design parameters and performance, the design principle for the bistable hybrid soft bending actuator (BH-SBA) of the cheetah-inspired high-speed crawlers is first studied. Fig. 1B shows the schematic design of the BH-SBA. It consists of a bistable flexible spine mechanism with a pre-tensioned linear spring and two identical soft bi-directional bending actuators as "skeletal muscle". Each soft actuator has two layers of embedded pneumatic channels (Fig. S1A) for actuating reversible bistability to generate switchable two-way bending upon pressurization control in dual channels (Fig. S1B). The 3D printed bistable spine mechanism is composed of two rigid hinged linkages to guide the bending motion of soft actuators (Fig. S1B). The pre-tensioned spring that connects two ends of the bistable mechanism can store and release potential mechanical energy. The geometrical and materials properties parameters of a BH-SBA prototype are listed in Table S1. As schematically illustrated in Fig. 1D, the working principle follows four steps. First, the spring is pre-tensioned (Fig. 1D(i)) and attached to both ends of the straight linkages (Fig. 1D(ii)) to enable a BH-SBA. The linkage-based spine with attached stress-free soft bending actuators and embedded pre-tensioned spring results in an unstable state (Fig. 1D(ii)), which possess the maximum energy as shown in the schematic energy profile of Fig. 1B (State I) and a zero bending angle $\theta$ ($\theta$ is defined in the inset of Fig. 1B). Second, after releasing the pretension in the spring, the BH-SBA bends and eventually rests in one equilibrium (i.e. the stable state shown in Fig. 1C(iii)), which possess local minimum energy with $\theta = -\theta_{eq}$ (State II in Fig. 1C). Third, upon pneumatic actuation, i.e. inflating the exterior layered pneumatic channel of the bending actuator at rest states ($\Delta P_1 > 0$ in Fig. 1D(iii)), the BH-SBA bends toward the unstable state, where the spring stores the most energy. Fourth, when it bypasses the unstable state, it rapidly snaps to the other stable state with $\theta = \theta_{eq}$ (Fig. 1D(iv) and State III in Fig. 1C). This bistable mode can be reversibly switched by following path (a) in Fig. 1C through actuating the other side of the channeled layer in the bending actuators. Thus, the actuator can generate a swing motion under dual channel actuation between two stable states. In the working regime of path (b) with the monostable mode, in step 3, the interior layered pneumatic channel ($\Delta P_2 > 0$ in Fig. 1D(iii)) is pressurized instead to further bend away from both unstable and equilibrium states for potential applications in soft manipulator discussed later, thus in this case bistability is not activated and only single channel actuation is needed. In the following, we will explore the mechanical behavior of the BH-SBA under both non-actuated and actuated states.

The competition between energy release and storage during bending of the soft pneumatic actuators and stretching of the spring determines the bistability of the system. Fig. 2A shows the analytical energy landscape of the bistable actuator without pressurization as a function of the bending angle $\theta$. The total potential energy $U_{total}$ (blue curve) is the sum of the strain energy in the soft bending actuator $U_{actuator}$ (mainly dominated by its bending energy, red curve) and the stretching energy in the spring $U_{spring}$ (yellow curve) (see details in Supplementary Materials). i.e.

$$U_{total} = U_{actuator} + U_{spring} \qquad (1)$$

$U_{actuator}$ shows a convex parabolic shape with respect to the bending angle with a minimum value of 0 J at $\theta = 0°$, whereas $U_{spring}$ exhibits a concave parabolic relationship with respect to the bending angle with a maximum energy of $U_I = k\Delta x_I^2/2$ at $\theta = 0°$, where $k$ is the spring stiffness and $\Delta x_I$ is the spring pretension length. The competition between $U_{actuator}$ and $U_{spring}$ leads to a symmetric profile of the total potential energy with one peak (unstable State I) at $\theta = 0°$ and two minimum points at two equilibrium and stable states II and III at $\theta = \pm \theta_{eq}$. The value of $\theta_{eq}$ can be obtained by minimizing the total potential energy (see details in Supplementary Materials), i.e.,

$$\frac{dU_{total}}{d\theta} = 0 \tag{2}$$

Fig. 2B shows the analytical equilibrium bending angle $\theta_{eq}$ from Eq. (2) as a function of spring stiffness $k$ and pretension length $\Delta x_I$. It shows that for all springs, $\theta_{eq}$ increases monotonically and nonlinearly with both $\Delta x_I$ and $k$, i.e., the more energy it stores in the pre-tensioned spring, the larger angle it bends after pretension release, which is consistent with the experimental data. However, it should be noted that the BH-SBA does not bend (i.e., $\theta_{eq}= 0$) upon release of the pretension until the spring is extended beyond a threshold value. It implies that the initial potential energy $U_I$ stored in the pre-tensioned spring should be large enough to overcome the strain energy in the soft actuator to trigger the onset of instability. The energy difference between maximum and minimum in Fig. 2A defines the energy barrier $\Delta E$ of the bistable system. It requires $\Delta E > 0$ to enable the bistability. Fig. 2C shows that the energy barrier $\Delta E$ can be tuned by either $\Delta x_I$ or $k$ in the spring (see details in Supplementary Materials), where $\Delta E$ increases nonlinearly and monotonically with both $\Delta x_I$ and $k$. Similarly, Fig. 2D shows that at a given spring constant $k$ (e.g. $k = 1.29$ N/mm), the stiffness $K$ of end-effector of the bistable hybrid bending actuator varies with the bending angle $\theta$ and increases dramatically with $\Delta x_I$. Here $K$ is defined as (see details in Supplementary Materials)

$$K = \frac{df_b}{d\delta}, \quad f_b = \frac{T}{0.5L\cos\theta}, \quad T = \frac{dU_{total}}{d\theta} \tag{3}$$

where $f_b$ is the static blocking force or reaction force at a given deflection of $\delta$, $T$ is the joint torque, and $L$ is the length of the hinged linkages. Fig. 2D shows that its stiffness can be manipulated over a large range from about 10 *N/m* to over 500 *N/m* by simply tuning $\Delta x_I$ from about 9 to 17 mm. The wide-range stiffness modulation provides the bistable actuator as a potential manipulator with tunable strength from low to high force as discussed later.

The switch between two equilibrium states in the BH-SBA can be actuated through pneumatic pressurization of the soft bending actuator to overcome the energy barrier. Fig. 2E shows the measured bending angle $\theta$ as a function of the pneumatic pressure $p$ for different spring pretension length $\Delta x_I$ when actuated from one stable state with a negative bending angle. For all springs, it shows a similar nonlinear J-shaped curve, where $\theta$ first increases gradually and monotonically with $p$ (the actuator unbends), followed by an instantaneous and steep rise when approaching to the unstable state (the actuator snaps to the other stable state instantaneously), indicating the onset of snap-through bistability. The steep increase of the bending angle in Fig. 2E also defines the critical pressure $P_c$ for actuating the bistability.

The highly nonlinear relationship between $p$ and $\theta$ shown in Fig. 2E poses challenges to control the bending angle of the BH-SBA with pneumatic pressure. However, it is feasible to constrain its maximum bending angle during cyclic motion, which is of important implications in locomotion when bistable actuators are integrated for the design of mobile robots discussed later. Here we set the angular position limit of the bistable mechanism to stop its rotational movement at a preset stopping angles $\theta_s$ (inset of Fig. 1C(iii)). By setting the value of stopping angles to be smaller than the equilibrium bending angle $\theta_{eq}$ predicted by Eq. (2), i.e. $\theta_s < \theta_{eq}$ (Fig. 1D), the maximum bending angle during swing motion can be constrained and is equal to be the preset stopping angle even when it is over pressurized.

Next, we examine the critical actuation pressure $P_c$ to enable the bistability of the BH-SBA with a preset stopping angle $\theta_s$. Here we focus on how the energy barrier determined by the initially stored energy in the spring affects $P_c$ by manipulating the pretension length $\Delta x_I$ ($k = 1.29$ N/mm and $\theta_s = 60°$ are fixed). It shows that $P_c$ increases nonlinearly with $\Delta x_I$. When $\Delta x_I$ is small (i.e., $\Delta x_I < \sim 5.5$ mm, highlighted by yellow color), $P_c$ increases linearly with $\Delta x_I$ at a small slope and the actuator rests at the equilibrium position after pretension release without arriving at the preset stop angle due to $\theta_s > \theta_{eq}$ (Fig. 2F). As $\Delta x_I$ further increases ($\sim 5.5$ mm $< \Delta x_I < \sim 8.8$ mm, highlighted by green color), $P_c$ increases dramatically and nonlinearly with $\Delta x_I$ and the actuator rests at the preset angle stopper after pretension release due to $\theta_s < \theta_{eq}$ (Fig. 2B). Consequently, it requires much higher pressure to actuate the BH-SBA. For example, for the actuator with $\Delta x_I \approx 5.6$ mm, it only requires $\sim 10$ kPa pressure to induce the snap-through. However, when $\Delta x_I$ is increased to $\Delta x_I \approx 8.6$ mm, a 1.5 times increase in $\Delta x_I$ leads to over 3.4 times increase in the actuation pressure ($\sim 34$ kPa) for the BH-SBA to snap. When $\Delta x_I$ further increases to be larger than 8.8 mm (highlighted by purple color), we observe that it fails to trigger the onset of snap-through bistability since the pneumatic actuation cannot overcome the high energy barrier even when we increase the input pressure to be as high as 100 kPa. There are several factors attributed to the high nonlinearity and the absence of snap-through bistability: (1) large input pressure always leads to the over-inflation of the air channel along the out-of-plane direction, which does not contribute to the in-plane-expansion induced bilayer bending. (2) the soft bending actuators in the bistable hybrid actuator always possess a non-uniform curvature, especially when preloaded with large spring pretension (i.e., with large initial bending angles). In this way, the middle section in the soft actuators always has a much larger curvature than sections at two ends. This large curvature may collapse or even close the air channel in the middle, making it difficult to be actuated even when the input pressure is significantly increased, which explains the region of no snap-through bistability shown in Fig. 2F.

**Bistability for Force Amplification and Rapid Response**

To understand the force amplification and rapid response effects enabled by bistability, we further study the actuation performances of the bistable hybrid actuator (BH-SBA), including motion (Fig. 3A, Movie S1), dynamic blocking force (Fig. 3B-3C), and response time (Fig. 3D), in comparison with its two counterparts with the same geometry after removing the spring to disable the bistability. One is the hybrid soft bending actuator (H-SBA) without spring (left inset of Fig. 3B, denoted as "hybrid"), and the other one is the soft bending actuator (SBA) with neither linkages nor springs (middle inset of Fig. 3B, denoted as "soft"). In the bistable actuator, the spring stiffness $k$ is 1.29 N/mm and the pretension in the spring $\Delta x_I$ is 6mm, which corresponds to an analytical equilibrium bending angle of $\theta_{eq} = 63°$ (Fig. 2E). To constrain its actuated swing angle right at the preset stop angle, we set $\theta_s = 60° < \theta_{eq}$ in the spine mechanism guided by the design principle discussed above. To enable the occurrence of snap-through bistability, we set the actuation pressure to be 20 kPa, which is moderately higher than the critical actuation pressure (Fig. 2F). The three actuators with their base structure fixed are pneumatically actuated at the same pressure of 20 kPa and the same frequency of 3.2 Hz during the swing motion (the timing control parameters are shown in Fig. S2 and Table. S2, and the experimental setup for measurement is shown in Fig. S3).

Fig. 3A shows the trajectories of deformed bodies and the maximum bending angles $\theta_{max}$ of the three bending actuators at rest states. Compared to its soft ($\theta_{max} \sim 55°$, bottom right of Fig. 3A) and hybrid ($\theta_{max} \sim 25°$, middle right of Fig. 3A) counterparts, the BH-SBA exhibits the largest $\theta_{max}$ (top right of Fig. 3A) and rests right at the preset stopping angle 60º as designed (Movie S2). The SBA possesses the largest deflection, resulting from both axial elongation and bending deformation. However, the axial elongation in both hybrid bending actuators is largely constrained by the rigid spine mechanism with a dominated bending deformation, which facilitates activating the bistability. In addition, the soft bending actuators in both hybrid actuators exhibit similar deformation motion, which follows the rigid rotation of the upper linkage around the joint in the spine mechanism. Thus, our model is able to analyze the bending angle and curvature of the hybrid bending actuators. In contrast, it is more challenging to analyze the curvature and bending angle of the SBA due to the axial elongation discussed above.

Fig. 3B shows the corresponding measured dynamic blocking force $F_b$ as a function of the bending angle during the swing motion of the three actuators, where the dynamic blocking force is much higher than the measured static blocking force by quasi-static indentation of the hybrid soft prototype (Fig. S4) and the bistable flexible linkages alone (Fig. S5). The dynamic blocking force $F_b$ is defined as the dynamic impact force measured from the tip of the actuator when its swing motion actuated from its rest position is blocked by a rigid plate at a certain angle. The dynamic blocking force is studied because it represents the ground reaction force of legged robots to characterize the propulsion capability for high-speed locomotion. The experiment shows that as the actuator bends away from the center with $\theta = 0º$ to $\theta = \theta_{max}$ with the largest deflection at rest states, $F_b$ drops monotonically with the increase of the bending angle for all the three actuators, and the BH-SBA demonstrates the highest blocking force. As expected, the SBA generates a smaller dynamic blocking force of less than 1 N due to its materials compliance. Surprisingly, despite its significantly enhanced stiffness with a hybrid design, the measured dynamic blocking force in the H-SBA is slightly lower than the SBA due to its much smaller swing angle $\theta_{max}$, where $\theta_{max} \sim 25º$ in the hybrid one is less than half of the swing angle $\theta_{max} \sim 55º$ in the soft one. In distinct contrast, the peak dynamic blocking force near the center with $\theta \sim 10º$ is significantly enhanced to $\sim 2.7$ N in the BH-SBA, which is approximately 3 times larger than the soft one and 7 times larger than the hybrid one. In addition, at the rest states with $\theta = \theta_{max}$, $F_b$ drops to approximately zero for both non-bistable actuators, whereas the reduced $F_b$ at the stop angle is still close to 2N for the BH-SBA, demonstrating the benefit of bistability in amplifying the output force capability. We further examine how the input pressure influences the dynamic blocking force $F_b$ of the BH-SBA at a given certain bending angle. The results in Fig. S6 show that $F_b$ increases monotonically first with the pressure input and then approaches a plateau, exhibiting a highly nonlinear behavior.

With the understanding of force amplification through bistability mechanism, we further exploit the enhancement of the dynamic blocking force by manipulating the initial mechanical energy storage in the spring, i.e. $U_I = k\Delta x_I^2/2$ through tuning pretension length $\Delta x_I$. We test three prototypes of the BH-SBAs with the same geometry but different spring pretension ($\Delta x_I = 6$ mm, 7 mm and 8 mm). All actuators are pressurized at 20 kPa to enable the bistability and the spring stiffness is kept as the same $k = 1.29$ N/mm. Fig. 3C shows the comparison of the measured dynamic blocking forces vs. bending angle at different spring pretension lengths. As expected, the larger spring potential energy it stores, the higher dynamic blocking force it generates. The actuator with $\Delta x_I = 8$ mm demonstrates the largest dynamic blocking force with a range of 2.5 – 3.4 N, which is about 30% higher than that with $\Delta x_I = 6$ mm. Meanwhile, as discussed above, an increased $\Delta x_I$ leads to a dramatic increased critical actuation pressure to trigger the occurrence of bistability, thus a relatively high pneumatic pressure is normally required to achieve a high blocking force.

In addition to maximum bending angle constraint and output force amplification, the proposed bistable hybrid system also contributes to rapid response during the swing motion. Upon the same actuation pressure of 30 kPa and the same air flow rate of ~ 3 L/min, we compare the response time of 6 prototypes with bistability being disabled or enabled in Fig. 3D and Table. S3. It shows that the H-SBA exhibits the longest response time 2.60 s to bend from $0°$ to its maximum bending angle $25.3°$ at 30 kPa, which is over twice of the time of 1.26 s for the SBA bent from $0°$ to its maximum bending angle ~$60°$. On the contrary, the BH-SBAs take the least response time of less than 1 s to switch from the far left (negative $\theta$) to the far right (positive $\theta$). Their response time is composed of two parts including before ($\Delta t_{II-I}$, highlighted by blue color in Fig. 3D) and after snapping-through ($\Delta t_{I-III}$, highlighted by pink color in Fig. 3D), i.e. bending from State II (negative resting angle) to State I (zero bending angle), and from State I to State III (positive resting angle), respectively. We find that as the spring pretension $\Delta x_I$ increases, $\Delta t_{II-I}$ increases dramatically from 0.26 s to 0.98 s to overcome the increased energy barrier. However, when beyond the unstable state, the snapping-through time $\Delta t_{I-III}$ decreases with increased spring pretension $\Delta x_I$ to release the increased stored energy more quickly. It is observed that for the studied prototypes, $\Delta t_{I-III}$ is less than 60 ms to snap to its resting state and the normalized snapping-through time, i.e. $\Delta t_{I-III} / \Delta t_{II-I}$ decreases dramatically from 0.24 to 0.05 when $\Delta x_I$ increases from 3.1 mm to 8.0 mm.

It is observed that the snap-through bistability in the BH-SBAs is well captured by the corresponding finite element method (FEM) simulations using ABAQUS (Simulia, Dassault Systems, 6.14) (Fig. 3E and Movie S3, see details in Supplementary Materials), where all actuators are actuated under a 40 kPa pressure and their transient dynamic responses with respect to time are recorded by the simulations. Despite the discrepancy between the absolute value of actuation and snap-through time in simulations with respect to experiments due to failing to modeling the pneumatic flow rate in experiments, the FEM results capture the trend of increased actuation time and decreased snapping-through time with the increase of spring pretension (Table. S4 and Fig. S7). Fig. S7 shows that for BH-SBAs with different spring pretension length, the relationship between simulated bending angle and actuation time exhibits *J*-shaped-like high nonlinearity, where the slope increases dramatically after the unstable state.

## 2. Bistability for High-Speed Crawler

Bioinspired by the flexion and extension motion of the spine in mammals such as the cheetah during the high-speed galloping (Fig. 4A), we explore harnessing the bistable spine mechanism enabled power amplification and fast response for its application in high-speed locomotive soft robots. As schematically illustrated in Fig. 4A, simply attaching four claws to the BH-SBA constitutes the proposed high-speed bistable hybrid soft robot. Similar to the active spine of the cheetah, the bistable linkage-based spine in the crawler bend downward to store energy and shorten the length of stride, and then bend upward bypassing the unstable straight configuration to quickly release the stored energy to increase its stride length. The active flexion and extension motion is driven by the soft pneumatic bending actuator as "muscles". The prototype is 7 cm long and 6 cm wide with a mass of 45 g. The spring stiffness $k$ is 1.29 N/mm and the pretension length is 6 mm at state I. The stopping angle of the spine is $\theta_s = 60°$. Small elastomer pads are attached on the claws, as shown in Fig. 4B, to break the deformation symmetry through the passive switchable friction for front and rear claws (*24*). At the rest state, its soft body is initially bent downward (top of Fig. 4B), the forefeet have much larger static friction force (because the Ecoflex elastomer is in contact with the surface) than the rear ones (right of Fig. 4B). This generates a forward motion when the soft body is actuated and bent upward; when the crawler is in a convex shape, i.e., bend-up (bottom of Fig. 4B), the rear legs have larger friction force than the front ones. This also results in forwarding locomotion when the body is actuated and bent downward.

Despite the simplicity of the design, the proposed bistable hybrid soft crawler exhibits a superior high locomotion speed (experimental setup in Fig. S8). It can achieve a linear locomotion speed of 174.4 mm/s, or 2.49 BL/s when pressurized at 20 kPa (average actuation frequency is 3.2 Hz, Fig. S2 and Table. S1) on a wooden surface (bottom of Fig. 4C and inset of Fig. 4F). Such a high locomotion speed is 2.1 times the speed of its counterpart soft crawler based on the SBA (middle of Fig. 4C and inset of Fig. 4F), and 4.7 times the speed of the hybrid soft crawler based on the H-SBA (top of Fig. 4C and inset of Fig. 4F). The comparison of real-time locomotion videos can be found in Movie S4. The slow motion of the bistable hybrid soft crawler (Movie S5) shows that its locomotion gait is similar to that of high-speed cheetahs during their energy-economic galloping (Fig. 4A). The forefeet touch the land bending up its back to store energy before they lift off, and then straightened and bent downward to lengthen the stride by extending the forefeet with all feet off the ground. In contrast, for the other two soft and hybrid counterparts, all the feet always remain in contact with the ground and thus consume more energy to overcome the friction (Movie S4). Note that different from the power supply from muscles and tendons in legs and spines of animals, the energy storage and quick release in the bistable hybrid soft crawler are triggered by the snap-through bistability in the sole spine mechanism, which saves energy cost and enables large force exertion and fast response for high-speed locomotion.

One of the advantages of this proposed technique is that we can change the speed of locomotion by tuning the energy barrier of the BH-SBA through spring pretension or spring stiffness. Next, we explore how the spring's stored energy due to spring pretensions affects the locomotion speed of the proposed crawler. According to Fig. 3C, a larger pretension in the spring, i.e., larger energy barrier, leads to a higher dynamic blocking force and thus higher reaction force in the bistable actuator. This improved force output contributes to a higher ground reaction force of the crawler's feet that provide the crawler with larger friction force to move forward. Thus, larger spring pretension length results in faster locomotion, which is verified by the experimental test shown in Fig. 4D and Movie S6. We compare the locomotion speed of three identical bistable crawlers but only with different $\Delta x_I$ = 6 mm, 7 mm and 8 mm in the spring, respectively. All crawlers are pressurized at 30 kPa with an average actuation frequency of 2.63 Hz (Fig. S2 and Table. S1). The result shows that the crawler with the largest spring pretension length ($\Delta x_I$ = 8 mm) demonstrates the highest locomotion velocity by leveraging high energy storage and rapid energy release. It can achieve a linear locomotion speed of 187.5 mm/s, or 2.68 BL/s, which is 18.7% faster than the crawler with $\Delta x_I$ = 7 mm and 38.8% faster than the crawler with $\Delta x_I$ = 6 mm. We further test its locomotion capability on a tilted surface. We find that when the surface is tilted to 17°, the bistable hybrid soft crawler can still achieve a fast location speed of 0.56 BL/s (Fig. 4E). However, the other two counterparts are not capable of climbing up the tilted surface (Movie S7).

In Fig. 4F, we compare the locomotion velocity of the proposed bistable hybrid soft crawler with a few representative locomotive soft and hybrid soft robotics (*7, 24-33*) and categorize them in the chart of body length speed vs. actuation frequency. These soft robot-based crawlers, mostly possessing continuous compliant bodies and stable structures, demonstrate a relatively slow speed in the range of 0.02 – 0.5 BL/s due to either slow actuation speed or small force exertion of the composed soft actuator, where the speed approximately increases with the actuation frequency for the reported crawlers. Disparate from its counterparts, despite the similar key component of soft materials-based soft actuator, the proposed bistable hybrid soft crawler is much faster, which is over 2. 5 times of the high energy-density-based dielectric crawler. Meanwhile it requires less input pressure (it only requires 20 kPa pressurization for operation) due to the benefit of rigid skeleton and force amplifier, which is even comparable, in velocity, with some rigid robots and terrestrial animals (1 – 100 BL/s) (*24*) . It should be noted that despite the hybrid design, the motion of the proposed crawler is actively and directly driven by the continuous deformation of the soft bending actuators under pneumatic actuation, while the passive deformation in the spine is driven by the soft actuators, playing the role as a passive force and speed amplifier. The stored energy in spring, which can be tuned by changing spring stiffness or pretension, plays a dominant role in its energy barrier, thus affects its force output, velocity, and energy efficiency. Ideally, a larger energy-stored spring generates a larger energy gap, thus corresponding to a larger force exertion from the bistable hybrid system. However, meanwhile, the high energy spring increases the energy consumption of the bistable hybrid system because it requires more energy input to overcome the high energy barriers. It may lead to a drop in the frequency and velocity of the bistable hybrid system unless external energy is increased (e.g. increase the airflow rate). Therefore, a trade-off should be considered for selecting the stiffness and pretension of the spring for the design of high-speed locomotive robots. Design optimization will be studied in the future to further improve the performance of the proposed high-speed crawler.

## 3. Bistability for High-Speed Underwater Swimmer

In addition to a terrestrial high-speed locomotive soft robot, the swing motion of soft body in fishes inspires us for exploring multifunctionality of bistable hybrid soft actuators as a high-speed underwater robot. The structure of this fish-inspired swimming robot is depicted in Fig. 5A. It is composed of an encapsulated bistable hybrid actuator and an attached thin plastic sheet-based fin (thickness = 0.25 mm) at its rear for enhancing the propulsion force. Compared to terrestrial locomotion, underwater locomotion requires a relatively higher output force to overcome the water resistance for propulsion due to the fluid-structure interaction. Therefore, for pneumatic bending actuator-based swimmer, it needs more energy input through a high actuation pressure 150 kPa, which could lead to the structural failure of its soft body if not strengthened. To enhance the force output and structural reliability, we encapsulate the soft bending actuators with conformable, stretchable, and bendable polymeric wrinkling-based envelope by following the method in Cianchetti et al. (*34*), as highlighted in the red color of wrinkled skin in Fig. 5A. When protected with braided sheath, the bending actuator can sustain higher pressure and survive longer fatigue-life (*34*); therefore, here we use the encapsulated soft bending actuator for underwater locomotion. The bending actuator is 45 mm in length and 25 mm in diameter. The whole prototype is ~150 mm long with a mass of 51g. The preset stopping angle of the linkages is $\theta_s = 45°$. The spring stiffness is 1.09 N/mm with 10 mm pretension length.

Similarly, we measure the dynamic blocking force and deformation of the encapsulated bistable hybrid actuator and compare with its two counterparts without bistability under the same pneumatic pressure of 160kPa. As expected, the bistable hybrid one generates the largest maximum dynamic blocking force over 4 N and maximum bending angle $\theta_{max}$ of 45°, which is equal to $\theta_s = 45°$. In contrast, its two hybrid and soft counterparts show smaller force exertion and smaller maximum bending angle (Fig. 5B).

Fig. 5C shows the comparison of swimming speed between the proposed bistable hybrid soft swimming robot and its two counterparts under the same pneumatic pressure of 160 kPa and average frequency of 1.3 Hz (Fig. S2 and Table. S1, experimental setup in Fig. S8): soft swimmer based on the encapsulated soft actuator and hybrid swimmer based on the encapsulated springless hybrid actuator. It shows that the bistable hybrid soft swimming robot can achieve a maximum average locomotion speed of ~117 mm/s (bottom of Fig. 5C), or 0.78 BL/s, which is 32% and 122% faster than its two counterparts: soft swimmer (middle of Fig. 5C) and hybrid swimmer (top of Fig. 5C), respectively. The comparison of real-time underwater locomotion between the three swimmers can be found in Movie S8.

We further compare the speed performance of our bistable hybrid soft swimmer with other reported soft swimming robots, which use different actuation methods (e.g. shape memory alloy and ionic polymer-metal composite actuators) and frequencies (Fig. 5D) (*35-40*). It shows that regardless of different actuation methods, the swimming velocities approximately follow a monotonically increasing trend with the actuation frequency, arriving at 0.7 BL/s at 5 Hz in the system of a manta ray-inspired electronic fish (*37*). In contrast, despite a relatively low average actuation frequency of 1.3 Hz, the proposed bistable hybrid soft swimmer achieves 0. 78 BL/s high speed (star-shaped symbol), outperforming most of the reported soft swimmers (round symbols). Although the bistable hybrid soft swimmer is still slower than biological fishes which typically exhibit a swimming capability of 2-10 BL/s (*41*), future work on optimization of the bistable structure, actuators, springs and the morphology of the robot hold great potential to fill this gap.

**4. High-Force Bistable Soft Gripper with Tunable Stiffness**

So far, we have demonstrated the versatility of bistability for force amplification and rapid response in enabling high-speed terrestrial and aquatic locomotive soft robots in the bistable operating regime. As schematically illustrated in Fig. 1D, bistability also provides another monostable operating regime in the actuator. Next, we explore leveraging the monostable regime for the design of high-force bistable soft grippers with wide-range stiffness modulation. The spring-based hybrid mechanism alone without activating bistability possesses the capability to produce high force for soft robots. The key of high-force output lies in the spring that supports a wide-range of tunable stiffness as demonstrated in Fig. 2D.

Stiffening/softening is an important feature of soft robots in maintaining their shape changes and realizing adaptable force exertion to different working environments. Research efforts have been dedicated to building soft robots with variable stiffness by harnessing granular jamming (*7*), phase changes (*42*), or tendon-driven-stiffening (*43*). But such robots have limited ranges of stiffness and limited force outputs. We employ a strategy similar to tendon-driven stiffening to our hybrid soft-rigid system. Despite the simplicity in implementation, we demonstrate its effectiveness in manipulating the stiffness as below.

Fig. 6A shows the schematic of the bistable hybrid soft gripper, which is similar to the encapsulated bistable hybrid actuator in Fig. 5A but operating in the monostable regime. The dual-actuation design allows independent operation of the pneumatic actuator (e.g. to handle delicate objects as Fig. 6D), or motor-driven tendon actuator (e.g. to handle stiff and heavy objects as Fig. 6E). A direct-current (DC) electric motor-driven tendon is attached to the end of the extension spring. When the DC motor is actuated, it pulls the tendon and thus extends the spring (Fig. 6C), bending the actuator to close the "fingers" towards the stopping angle. Further extension of the spring beyond the stopping angle significantly stiffens the actuator. For this tendon-spring-driven system, the spring stiffness plays a dominant role in determining the stiffening modulation. Thus, using a spring with higher stiffness ensures a larger amplification in stiffness. To understand this effect, we characterize its variable stiffness with static blocking force (N) as a function of the spring extension. The pincer grasping posture is shown in the inset of Fig. 6B. In the test, the actuator is bent at 80° (the stopping angle in the linkages is $\theta_s = 85°$) with its free end blocked by a force sensor. Simultaneously, the motor pulls the spring (the spring stiffness is 9.67 N/mm) and the bending actuators are pressurized at 120kPa. The experiment results (Fig. 6B) show that its static blocking force can be manipulated between 0.1 – 103 N by tuning the length of the spring, which agrees well with the simple theoretical model (see Supplementary Materials for details). As a soft variable actuator, it demonstrates a 1-$10^3$ range in stiffness modulation by harnessing the spring-tendon-stiffening system, which is difficult to achieve with most previously reported stiffening methods used for pneumatic actuators (*42, 44*). In addition, unlike stiffening mechanisms that only allow for binary (or limited) stiffness control (*45, 46*), the proposed stiffening strategy provides continuous tunability via tensioning of the spring, thus providing the potential to construct soft machines by harnessing a wide-range of stiffnesses modulation.

Taking advantage of this wide-range modulation and programmable stiffness, we design and fabricate a strength-adjustable hybrid soft gripper, which allows for manipulating a variety of objects ranging from fragile lightweight to high-load objects. As schematically shown in Fig. 6C, the gripper is composed of two encapsulated bistable hybrid actuators, each one is driven by the independently controlled pneumatics and DC motors (Fig. S9). Pneumatic actuation is used for gripping lightweight and fragile objects by utilizing the intrinsic compliance of soft materials/structures. This soft gripper requires the tendon in slack and thus the spring remains inactive, while motor-driven actuation is used for grasping heavy objects through pulling the spring to amplify the bending stiffness. Based on this simple working mechanism, we successfully demonstrate that the proposed hybrid soft gripper is versatile to grasp objects in a variety of shapes, sizes, and weights. For example, when pressured at 90kPa without activating the tendon (i.e., the tendon and spring remain relaxed), the gripper is capable of grasping a fragile egg, a small-size reagent bottle, a steel wrap, and tape rolls in regular spheroidal or cylindrical shapes as shown in Fig. 6D and Movie S9. When driven by motors, the gripper is capable of directly grasping much heavier objects such as a bottle of water (620g, right of Fig. 6D) and a 3.6 kg payload of dumbbells by wrapping around the objects (Movie S9). The grasping payload is further enhanced by pulling across the gripper's "finger" tips, for example, a 9.2 kg payload of dumbbells (Movie S9). Currently, the largest payload that our gripper can grasp is 11.4 kg (25.3lb), as demonstrated in Fig. 6E and Movie S9, which is ~180 times the weight of a single actuator. For all the tests, we use the same high-stiffness spring ($k = 9.67$ N/mm) and it is stretched by 3 mm at the stopping angle position for grasping the maximum payload. It should be noted that the grasping force of our gripper is mainly determined by the spring. Therefore, its grasping capacity can be further improved by using springs with higher stiffness. More details and grasping videos can be found in the Supplementary Materials.

## DISCUSSION

We demonstrate that by harnessing the rapid response of snap-through bistability, bistable hybrid soft machines can be customized for different applications with significantly improved mechanical performance, including high speed, high force, tunable bistability, and programmable stiffness. Compared to conventional stable soft bending actuators, bistability enables not only 20 times faster response time (tens of milliseconds) but also over 3 times higher exertion force. We develop an energy-based theoretical model to guide the design of bistable soft actuators, including the relationship among design parameters, control inputs, and outputs. This paper investigates two operating regimes of bistability for multifunctional high-performance soft robot design: bistable mode through the competition-synergy relationship between springs and pneumatic actuators, and monostable mode through tunable spring pretension and spring stiffness. It serves as a design guideline for building such soft actuators/robots. Although we only demonstrate the integration of our actuator in building high-speed locomotive robots and high-strength manipulators, the high-speed and high-force output without sacrificing the advantages of soft robots could find potential applications in surgical robots, wearable devices, and etc. The proposed bistable hybrid system opens a new avenue for constructing next-generation high-performance soft robots across multi-scales with multi-functionalities, and we envision that it could also inspire further exploration into building large-scale ($\geq 10^1$ cm) soft machines.

Despite the use of PneuNets-type conventional pneumatic soft actuators in our demonstrated hybrid systems, the proposed bistable strategy could be expanded to a broad range of soft robots and soft actuators (e.g. fluidic elastomer acutators, shape memory actuators, electroactive polymers, and stimuli-responsive materials) in response to different actuation inputs, including pressure, heat, electricity, light, and magnetic field. We believe that the proposed bistability strategy could also significantly improve their performance in force exertion, velocity, stiffness, and energy efficiency.

Inducing bistability in soft bending actuators potentially makes it more difficult to control the bending motion. In this work, we intentionally avoid controlling the bending angle after the onset of snap-through (Fig. 2E) to realize rapid motion. It might be challenging to control the bending angle to a desired value in a short time period because soft pneumatic actuator has low bandwidth. However, the controllability of the angle positions can still be enhanced by properly activating the antagonist pneumatic actuator, similar to how a conventional soft bending actuator is controlled (*5*), but considering the additional force generated by the spring. Another issue of the proposed hybrid system is that its overall degree of freedom is reduced since a rigid linkage with a revolute joint is used. This can be addressed by replacing the rigid linkage with a single-piece and compliant mechanism (e.g. the carbon fiber reinforced prepreg design of the flytrap-inspired bistable robot (*47*)) or pre-strained soft elastomers for energy storage (*32*). Thus, bistable soft robots made of entirely soft materials could be realized.

Future improvement of our hybrid soft machines includes: (1) increasing the energy storage density and peak power. The maximum stored energy in the spring is limited by the pneumatic actuation to overcome the high energy barrier. The current stored energy density of the bistable hybrid soft actuator is about 1.2 J/kg and the peak power is about 8 W (Table S1), which is far below the combustion-based jumping soft actuators (36). Improvements could be achieved through reinforcing the soft actuator or hydraulic actuation. (2) achieving active control of the spring pretension length during the motion of the bistable actuator, e.g., driven by DC motor or using shape memory spring. This active control will enable programmability in force, velocity, energy barrier, and stiffness of the bistable actuator, therefore allowing the soft robotic crawler or swimmer to change speed and achieve new locomotion modes such as jumping or escaping. (3) optimizing the proposed crawler and swimmer for faster locomotion. This could be achieved through structural optimization of the bistable actuator, replacing the rigid legs with energy-stored legs, or replacing the PneuNets with modified fast pneumatic networks for a higher actuation freuquency (48). (4) constructing a modular hybrid robot (with the bistable actuator as a single module) that enables multiple-degree-of-freedom (for example, by positioning multiple modules in orthogonal directions) and multi-stability for various applications. (5) improving sensing and control. Current robot prototypes use simple sensors (e.g., pressure sensors) to perform low-level control for the pneumatic actuator, and the Arduino microcontroller controls the timing of coordination between the pneumatic actuation and the bistable mechanism (Table. S2). In our future work, we will investigate sensing and control (e.g., how to use inertial measurement unit to measure joint angles to control the speed of locomotion of our robotic crawler) for improved performance and intelligence.

## MATERIALS AND METHODS

### Motion characterization

The motions of studied bending actuators (Fig. 3A) were captured by a high-speed camera (model MQ022CG-CM, USB3, Ximea GmbH) with a 60 Hz external synchronization signal. The recording time was ten seconds generating 600 frames. The tracking method was based on using the simple linear iterative clustering (SLIC) algorithm to generate trackable, meaningful super-pixels, followed by a 2D Kalman filter to predict the positions of actuators (49).

### Fabrication of bending actuators

The pneumatic soft bending actuators without encapsulation were fabricated following the conventional molding-demolding manufacturing technique for fluid-driven soft actuators reported by Ref. (1). Encapsulated pneumatic bending actuators were fabricated following Ref. (34). The braided structure (denoted by red color in Fig. 5A) was prepared by compressing the polyester sleeve (McMaster-Carr, 9284K4) through a steel rod and then curing in the oven at $60^o$ for 20 min. All rigid skeletons, mostly made of PLA, were 3D printed by Ultimaker 2+ and were bonded with the soft actuators by both Ecoflex 00-50 (Smooth-on Inc) and Insta-Cure glue (Bob Smith).

## Supplementary Materials

Text
Fig. S1. Structure of bistable hybrid soft bending actuator (BH-SBA).

Fig. S2. Actuation timing control for BH-SBAs
Fig. S3. Experimental setup for the measurement of bending angle, pressure, and flow rate for bistable actuator.
Fig. S4. Static blocking force and joint torque of BH-SBA.
Fig. S5. Force-displacement curves of the sole bistable linkages of BH-SBA under quasi-static indentation.
Fig. S6. The dynamic blocking force of BH-SBA vs. input pressure.
Fig. S7. The simulated bending angle vs. response time for BH-SBAs with different spring pretensions
Fig. S8. Experimental setup of soft robotics crawler (left) and soft robotic swimmer (right).
Fig. S9. Experimental setup of bending stiffness measurements.

Table. S1. List of geometrical and materials parameters for BH-SBA.
Table. S2. Data of actuation pressure and time control pattern.
Table. S3. The time elapse of SBA, H-SBA, and BH-SBAs with different spring pretensions.
Table. S4. The simulated time elapse of BH-SBAs with different spring pretensions.

Movie S1. Slow motion of swinging of bistable hybrid soft bending actuator (BH-SBA) captured by a high-speed camera.
Movie S2. Comparison of the real-time swing motion between BH-SBA and its two counterparts, the hybrid soft bending actuator (H-SBA), and soft bending actuator (SBA) with the same swing angle of 60º.
Movie S3. FEM simulation on actuating the bistability of BH-SBAs with different spring pretensions
Movie S4. Comparison of the real-time locomotion on a horizontal surface between the three crawlers based on integrated BH-SBA, H-SBA, and SBA.
Movie S5. Slow motion (x 0.125) of the BH-SBA-based crawler locomoting on a horizontal surface.
Movie S6. Comparison of the real-time locomotion of three BH-SBA-based crawlers on a horizontal surface with different spring pretension.
Movie S7. Comparison of the real-time locomotion on a slightly tilted surface between the three crawlers based on integrated BH-SBA, BH-SBA, and SBA.
Movie S8. Comparison of the real-time underwater locomotion between the three fish-like swimmers based on integrated BH-SBA, H-SBA, and SBA.
Movie S9. Demonstrations of strength-adjustable bistable hybrid soft grippers in grasping a variety of objects ranging from fragile lightweight to high-load objects.

**Funding:** J. Y. acknowledges the funding support from National Science Foundation (CMMI-1727792 and CAREER-**1846651**).

**Author contributions:** Y. T. and J. Y. developed the concept, designed the experiments, and analyzed the data. J. Y. supervised the project. Y. T. and Y. C. fabricated and characterized the prototypes. J. S. and J. Z. conducted the finite element simulation. O. H. M. and A. S. conducted the data analysis on tracking the swing motion of bending actuators. T. H. and H. S. conducted the model analysis and pressure-bending angle measurements. Y. T., J. Y., and H. S. wrote the manuscript. All the authors contributed to the discussion, data analysis, and edit of the manuscript.

**Competing Interests:** The authors declare that they have no competing interests.

**Data and materials availability:** All data needed to support the conclusions are presented in the main text or the Supplementary Materials. Additional data available from authors upon request.

# Figures and Tables

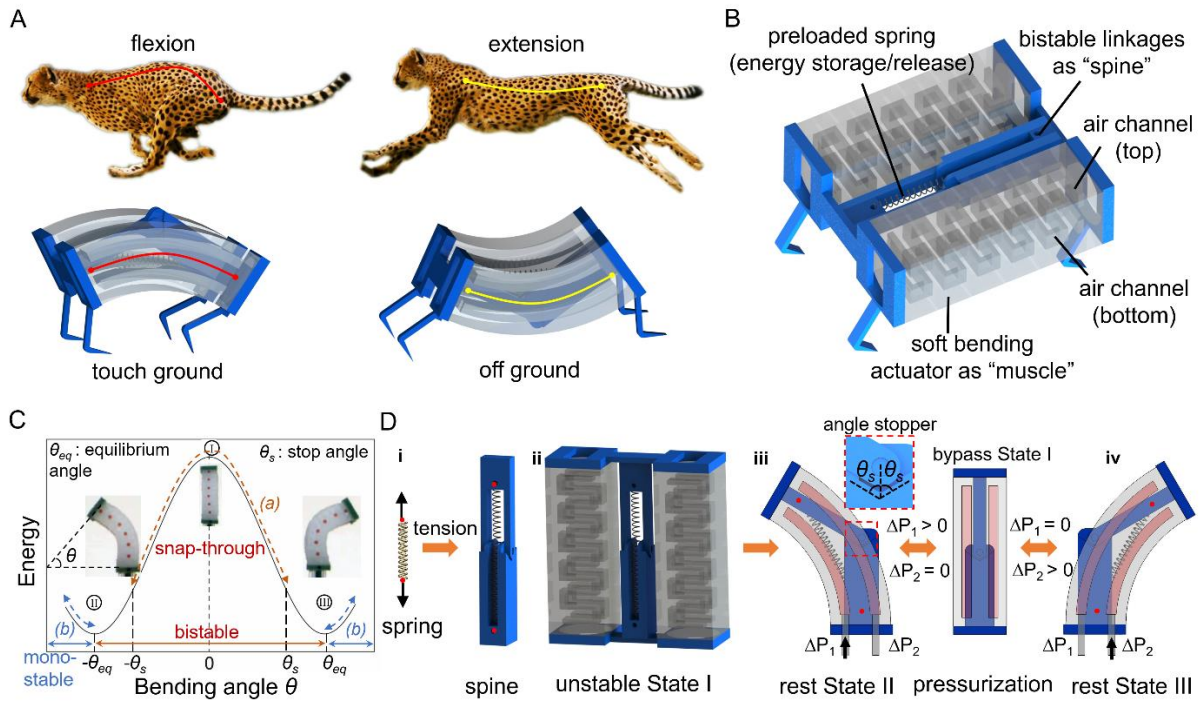

**Fig. 1. Spine-inspired bistable soft actuators.** (A) Bioinspired by the active spine mechanism during cheetahs' high-speed galloping, a bistable spine-based hybrid soft actuator is proposed to realize the similar spine flexion and extension through reversible snap-through bistability for design of high-speed locomotive soft robots. (B) Schematic design of a bistable hybrid soft bending actuator. It consists of three components: two soft pneumatic two-way bending actuators as skeletal muscle; a 3D printed flexible mechanism composed of two rigid hinged links as a spine; and a pre-tensioned spring that connects two ends of the mechanism for potential mechanical energy storage and release. (C) Schematics of energy landscape of the bistable actuator, showing one peak (unstable State I) and two localized minimum energy states (stable State II and III). It provides two operating regimes: one is the bistable switch in path (a) and the other is the monostable state in path (b). Insets show the corresponding ecoflex-based bistable actuator prototypes at each state. (D) Schematic illustration of the bistable working mechanism under both non-actuated (spring pretension release at resting states in i-iii) and actuated states (reversible snapping-through under pneumatic actuation in iii-iv). Inset shows the set of an angular stopper to constrain the maximum bending angle at the preset stopping angle of $\theta_s$.

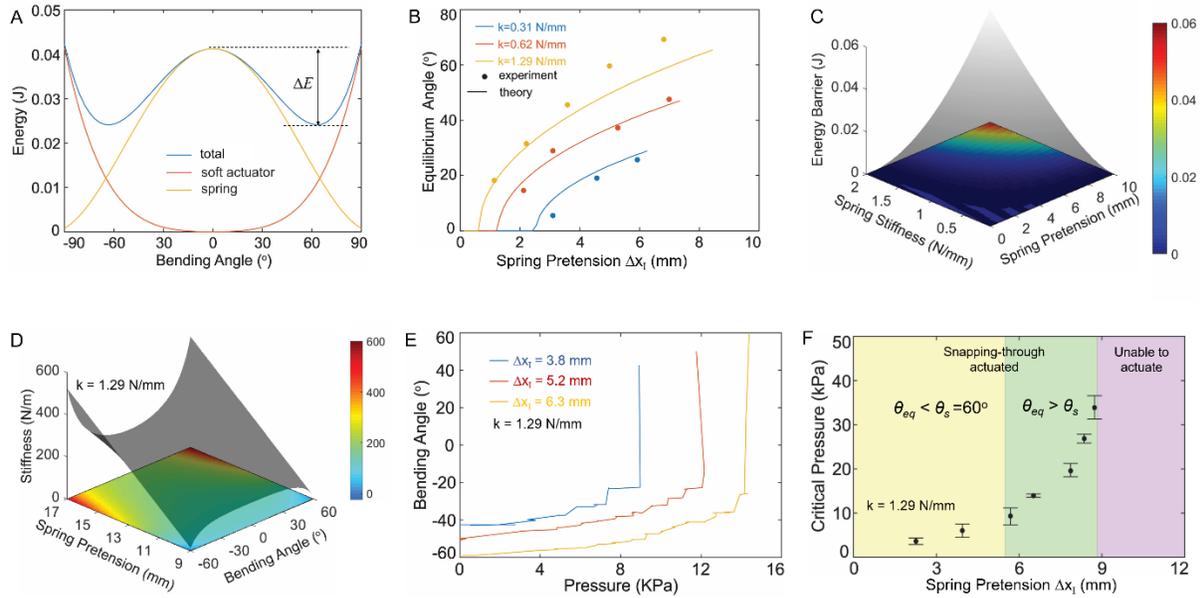

**Fig. 2. Design principle of energy competition and synergy relationship between springs and soft pneumatic actuators.** (A) The total energy of the bistable system (blue curve) as sum of strain energy in the soft bending actuator (red curve) and stretching energy in the spring (yellow curve). The energy difference between the peak and valley defines the energy barrier $\Delta E$. (B) Comparison between theory and experiments on equilibrium bending angle vs. spring pretension length at State I ($\Delta x_I$). (C) Energy barrier $\Delta E$ as a function of $\Delta x_I$ and $k$ in the spring. (D) Actuator stiffness as a function of spring pretension length and bending angle. (E) The measured actuated bended angle change with the increase of pressure $p$ during the snapping-through bistability switching from State II to State III for the bistable actuators with different pretension. (F) Critical pressure for pneumatically actuating the bistable switch between two stable states as a function of pretension length $\Delta x_I$. It exists a threshold value of $\Delta x_I$, below which bistable switch can be activated (yellow and green zone) while beyond which it fails to be activated (purple zone) due to the large energy barrier. The resting position after pretension release is at the equilibrium angle $\theta_{eq}$ with $\theta_{eq} < \theta_s = 60°$ in the yellow zone and at the stopping angle of $\theta_s = 60°$ in the green zone.

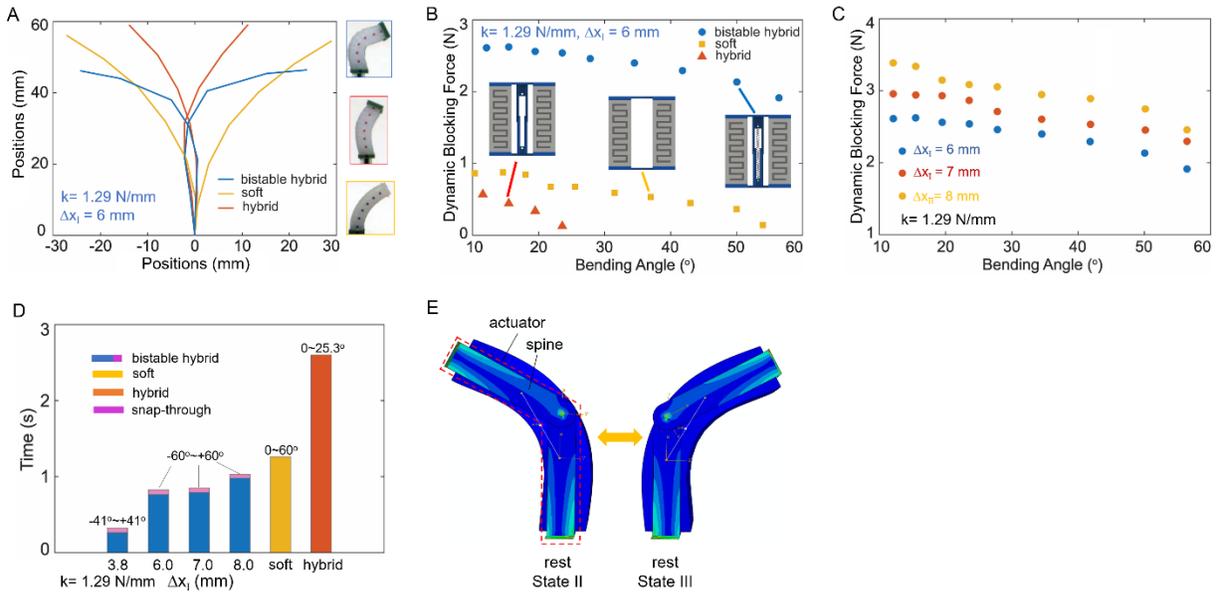

**Fig. 3. Bistability for amplified force and fast response.** Comparison of mechanical performances between the bistable hybrid actuator (blue color) and its two bistability-disabled counterparts: a soft actuator without both linkages and springs (yellow color) and a hybrid actuator without springs (red color). (A) Tracking of the bended shapes at two stable states. (B) Dynamic blocking forces at different blocked bending angles. Insets are schematics of front views of the three actuators. (C) Dynamic blocking force as a function of spring pretension. All the actuators in (A), (B), and (C) are pressurized at 20 kPa. (D) The response time of achieving the largest bending angle for three bistable, hybrid, and soft actuators with different spring pretensions under the same actuation. All actuators are pressurized at 30 kPa with the same flow rate of 3 L/min. For bistable ones, the response time is composed of two parts including before (highlighted by blue color) and after snapping-through (highlighted by pink color). (E) finite element simulation results on the actuated switch between two stable states of the bistable hybrid actuator through pressurization.

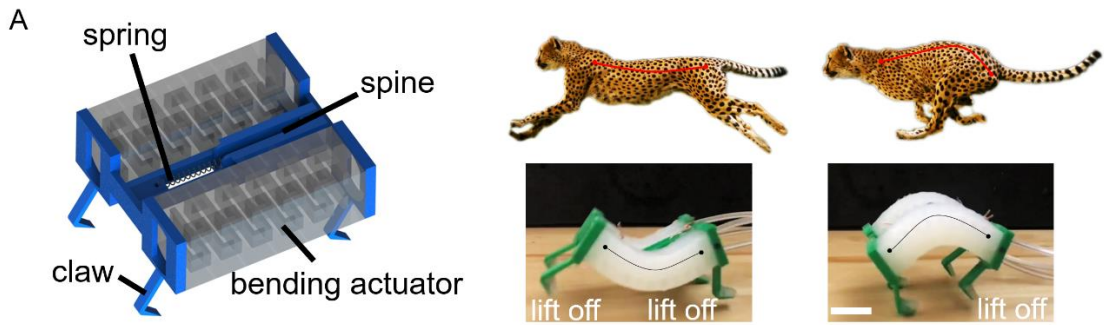

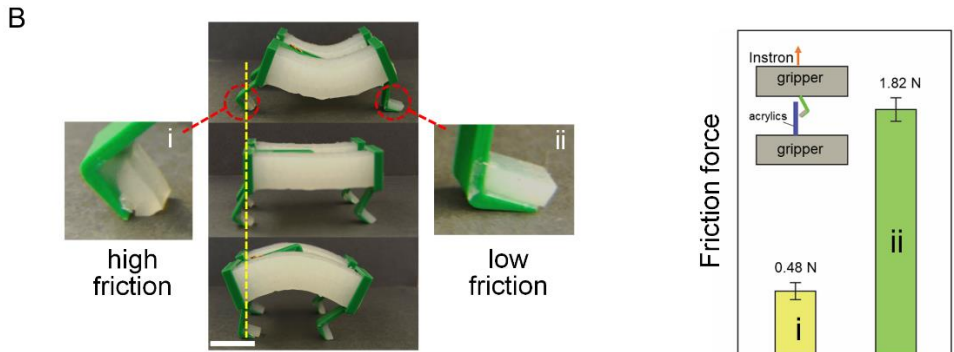

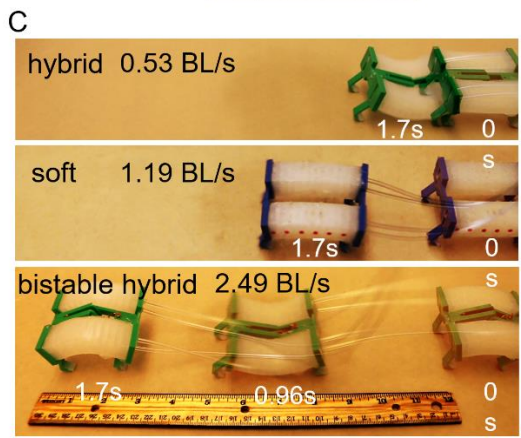

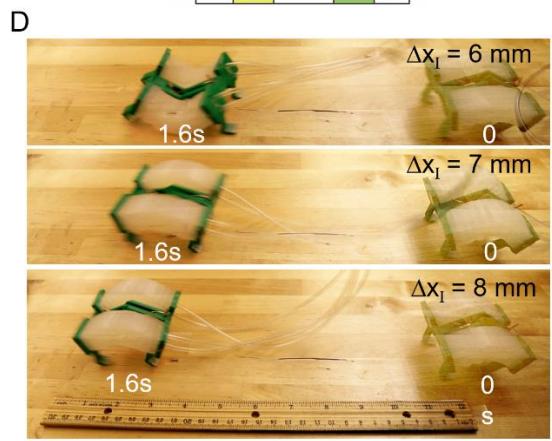

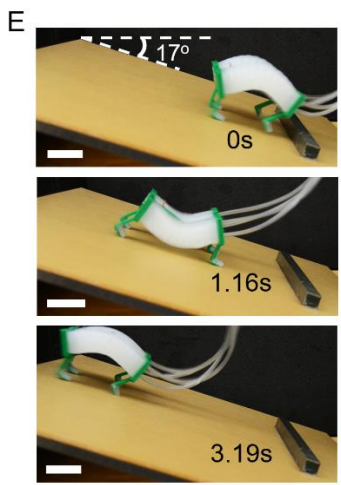

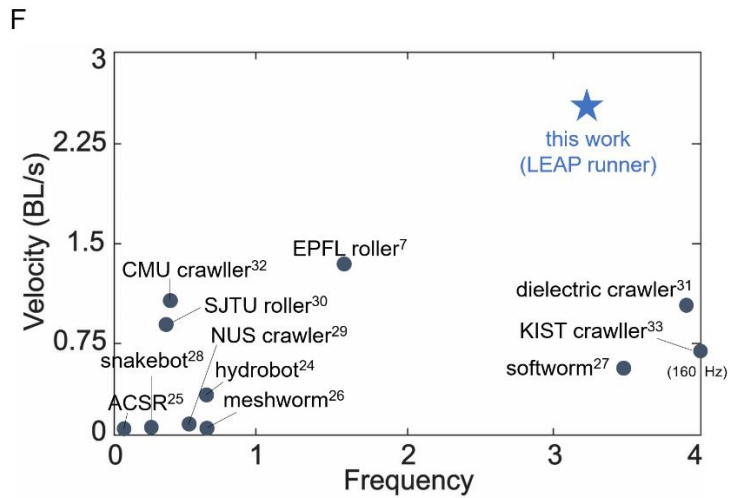

**Fig. 4. Bistability for high-speed crawler** (A) Locomotion gaits of the proposed bio-inspired crawler by the spine actuation in the fastest galloping cheetahs. The spine bends upward to store energy when touching ground and bends downward to release energy and extend its stride length with legs lifting off the ground during the high-speed locomotion. (B) Left: Mechanism of directional locomotion. The Ecoflex elastomer pads attached to the claws provide tunable friction force (i: high friction. ii: low friction) and transit the symmetric bending of the bistable actuator into directional locomotion. Right: static friction forces of claws with and without Ecoflex pad-substrate contact. The claw shows ~280% increase in the friction force (1.82 N) when the attached Ecoflex pad contacts the substrate. Inset shows the schematic of friction force measurement. (C) Demonstration of locomotion in the high-speed bistable crawler and its two counterparts with disabled bistability at different actuation time: soft crawler based on the soft actuator and hybrid soft crawler based on the hybrid one. All actuators are pressurized at the same pressure of 20 kPa and the same frequency of 3.2 Hz. The bistable hybrid soft crawler shows the fastest speed (~2.49 BL/s). (D) Demonstration of locomotion in the high-speed bistable crawler with different values of spring's stored energy in the bistable actuator. All actuators are pressurized at the same pressure of 30 kPa and the same frequency of 2.63 Hz. The one with largest spring pretension shows the fastest speed (~2.68 BL/s). (E) Demonstration of the proposed bistable hybrid soft crawler's capability in climbing slightly titled surfaces (tilting angle of 17º). The other two counterparts fail to climb. The scale bar is 25 mm. (F) Comparison of locomotion velocity measured in body length per second (BL/s) between our proposed bistabe hybrid soft crawler (denoted as star-shaped symbol) and reported locomotive soft robots in literatures.

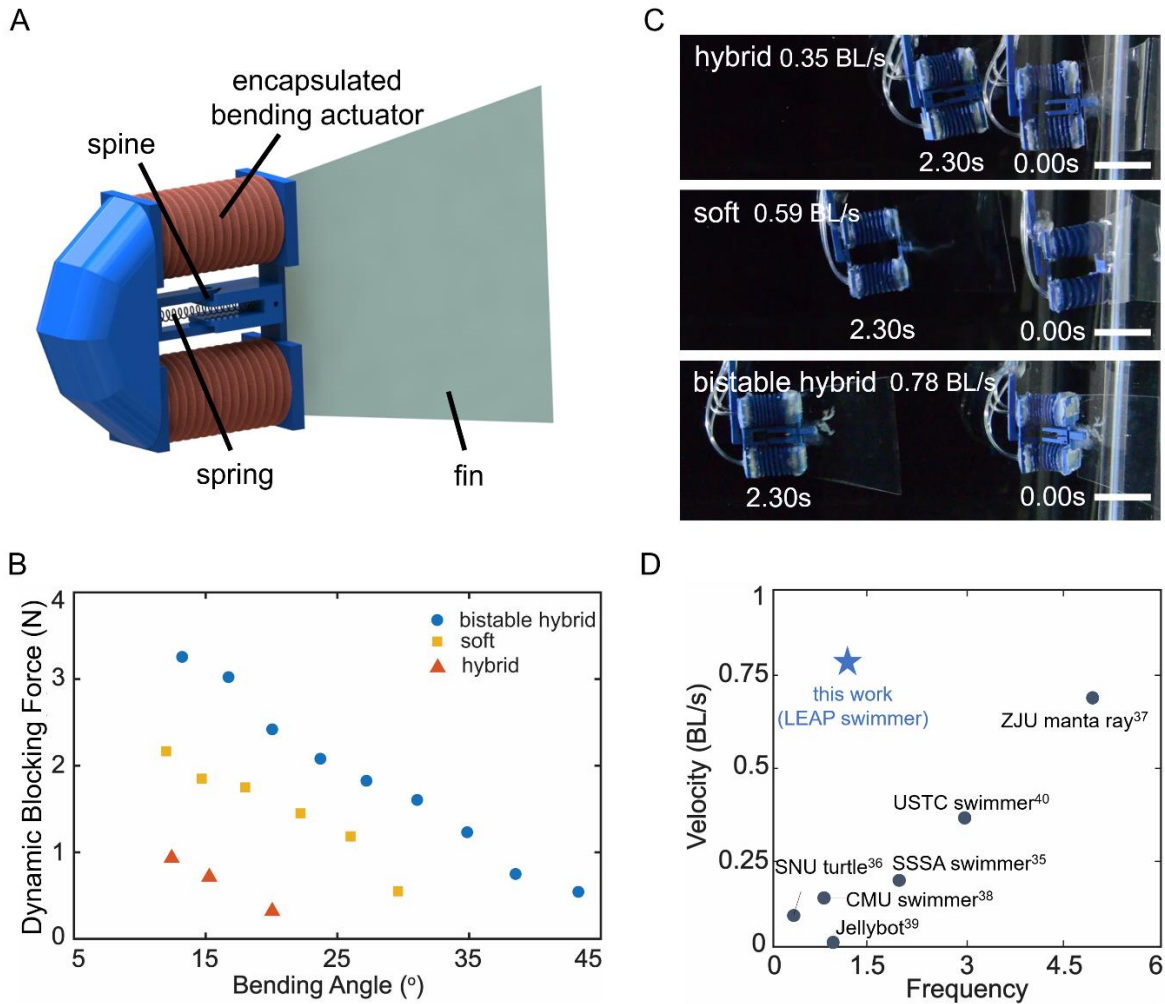

**Fig. 5. Bistability for high-speed underwater fish-like soft robot.** (A) Schematic of the proposed fish-like robot, which is composed of the bistable actuator attached with a polymeric fin. The schematic head is for decoration purpose only. (B) Comparison between the bistable actuator and its two counterparts of hybrid and soft actuators in dynamic blocking force vs. blocked bending angle. All actuators are pressurized at 160 kPa and an average frequency of 1.3 Hz. (C) Demonstration of underwater locomotion in bistable hybrid soft swimmer and its two counterparts at different actuation time: soft swimmer based on encapsulated soft actuator and hybrid soft swimmer based on encapsulated hybrid actuator. The bistable hybrid soft swimmer shows the fastest speed. The scale bar is 50 mm. (D) Comparison of swimming velocity between the proposed bistable hybrid soft fish-like robot (denoted as star-shaped symbol) and various reported underwater soft swimmers (denoted as round symbols).

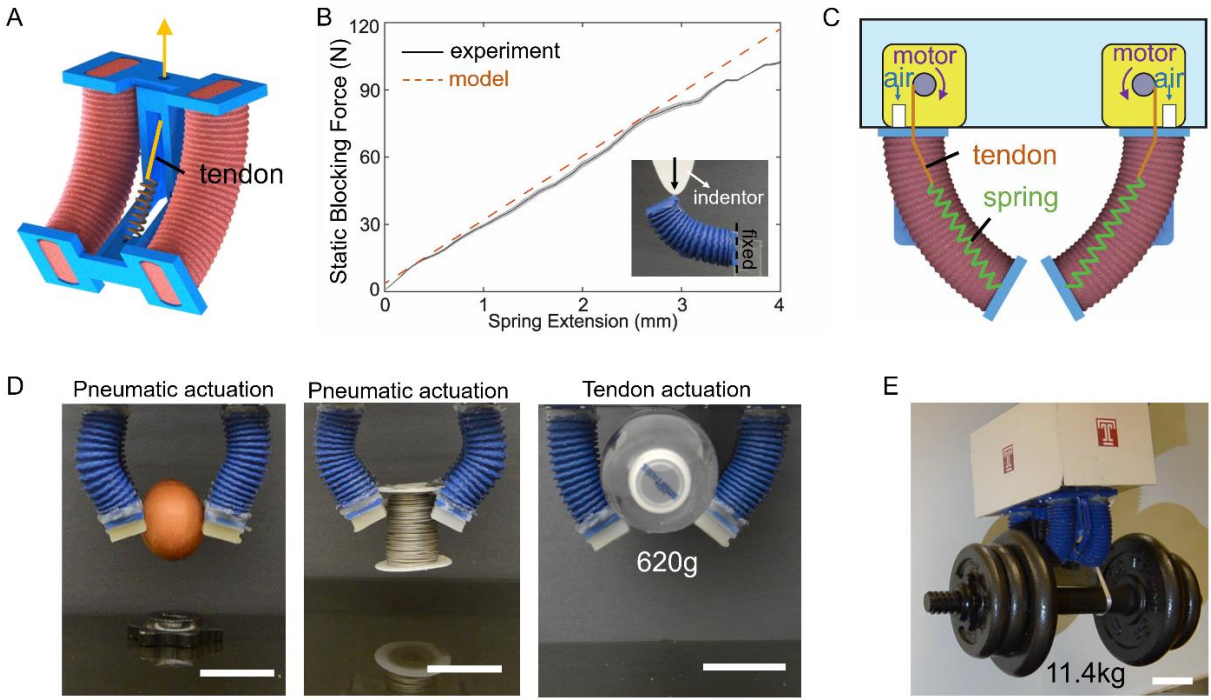

**Fig. 6. Bistable hybrid soft grippers with wide-range variable stiffness modulation.** (A) Schematic of the encapsulated bistable hybrid actuator driven by tendon. (B) Stiffness test of the encapsulated actuator (bent at 75º) (insets). We characterize the force as a function of spring extension. The solid line is the mean of three experiments (with shaded error bar) and the dashed line presents the theoretical model result. The stiffness is calculated as the slope of the force vs. indentation depth. (C) Schematic illustration of the proposed gripper by assembling two actuators. The manipulation can be controlled by both motors, through pulling the spring, and pneumatic signals. (D) Demonstrations of its capability in grasping various objects in different shapes and weight. (E) Demonstration of high-load manipulation. The scale bar is 50 mm.